\documentclass{article}
\usepackage{graphicx}

\begin{document}

\begin{center}
{\large \bf Seasonal Variations of the Effect from Solar Neutrinos
in Lithium Detector.}%

\vskip 0.3in

Anatoli Kopylov and Valeri Petukhov\\ Institute of Nuclear
Research of Russian Academy of Sciences \\ 117312 Moscow, Prospect
of 60th Anniversary of October Revolution 7A
\end{center}

\begin{abstract}
The presence of two monochromatic lines of approximately equal
intensity: $^{7}$Be- and pep-neutrinos in the sensitivity plot of
lithium detector makes the pattern of the seasonal variations of
the effect from solar neutrinos very characteristic in case if the
long-wave vacuum oscillations are realized. This can give the very
high accuracy in the measurement of the parameters of neutrino
oscillations especially if combined with the results obtained by
the detector sensitive mainly to $^{7}$Be-line like BOREXINO or
KamLAND.
\end{abstract}

    The recent results obtained by the SNO detector \cite{1}
compared with the previous results of SuperKamiokande \cite{2}
have shown that in the flux of solar neutrinos only approximately
1/3 are the electron neutrinos while the rest 2/3 are the
$\mu$-,$\tau$-neutrinos which proves that neutrinos do
oscillate.This result albeit needs further confirmation may have a
dramatic influence on the neutrino physics in general and on the
neutrino astrophysics in particular. The important issue of this
is the determination with the highest possible accuracy the
parameters of neutrino oscillations. Which way the further
development will take depends very much on $\Delta $m$^2$ realized
in the solar neutrino data. If LMA region with $\Delta $m$^2$ of
about 10$^{-5} \div 10^{-4}$ eV$^{2}$ is the solution this will
open great field of research for neutrino factories with the
fascinating perspectives to observe CP and T violations in
neutrino oscillations \cite{3}. But if the longwave vacuum
oscillations are the choice of nature then there will be a good
opportunity to observe the seasonal variations in solar neutrino
detectors. This possibility looks attractive both for electronic
\cite{4} and for radiochemical \cite{5} detectors, the only thing
which is crucial is the accuracy of measurements. It will be shown
in this paper that lithium radiochemical experiment is able to
furnish the valuable information on this subject and that the
combination of the results of lithium detector with the results of
the one sensitive mainly to $^{7}$Be neutrinos like BOREXINO
\cite{6} or KamLAND \cite{7} will enable to fix parameters of
neutrino oscillations with very high accuracy.

The lithium detector was proposed by Bahcall \cite{8} and is
utilizing the reaction of neutrino back capture on lithium with
the threshold of 0.86 MeV for the ground state to ground state
transition:

\begin{center}
$^7$Li + $\nu $ $\rightarrow $ $^7$Be + e$^{-}$
\end{center}

\noindent The ground-state to excited-state transition

\begin{center}
$^7$Li + $\nu $ $\rightarrow $ $^7$Be* + e$^{-}$
\end{center}

\noindent with the energy of the excited state 0.478 keV
contributes about 10\% to the total signal from solar neutrinos.
The cross-section of this reaction is relatively high and can be
computed with very high accuracy because both transitions are
superallowed \cite{9}. Another advantage of lithium target is that
in the sensitivity plot, as one can see on Fig.1, two
monochromatic lines:

\begin{figure}[!h]
\centering
\includegraphics[width=3in]{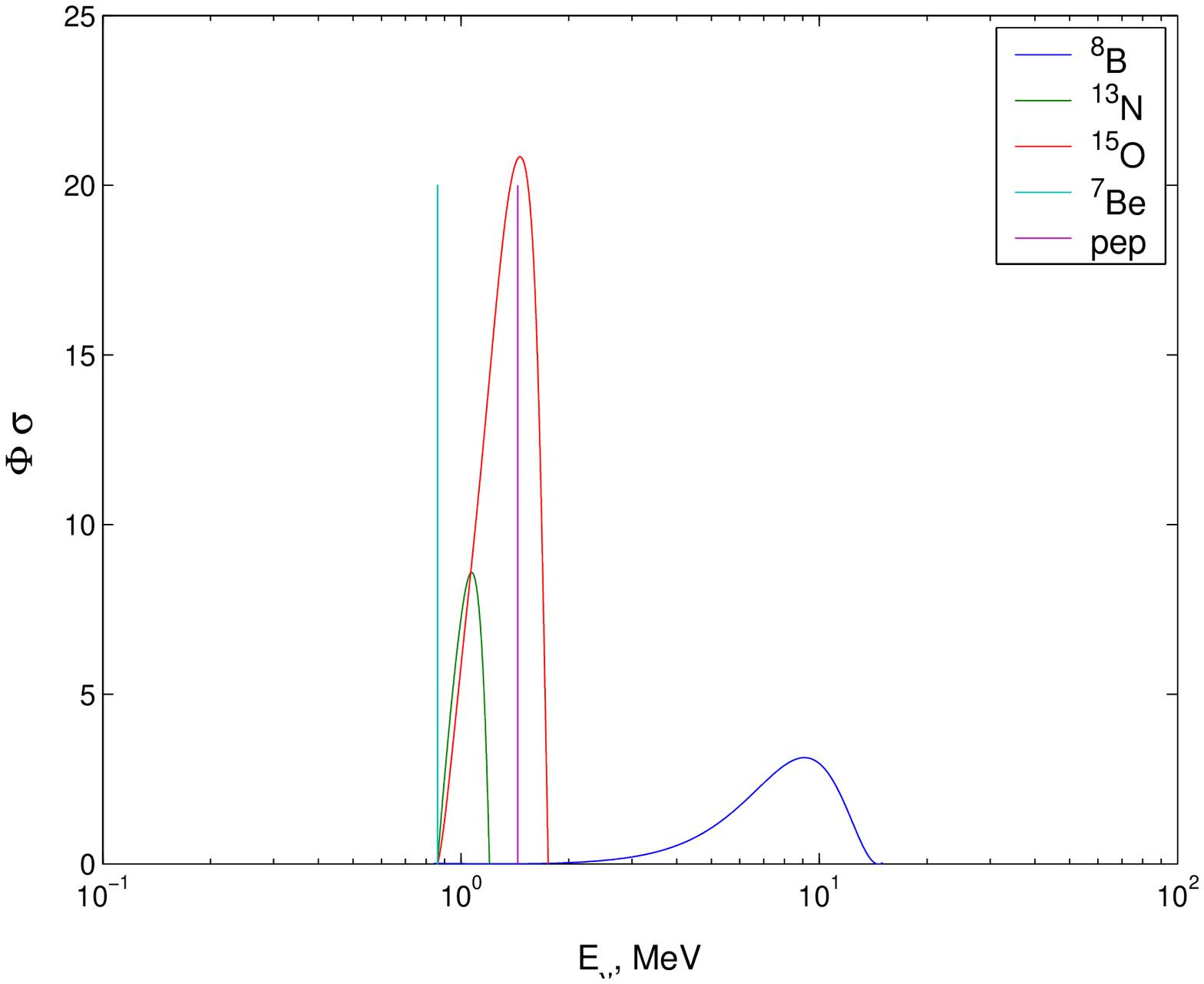}
\caption{The sensitivity plot of lithium detector}
\end{figure}

\noindent $^7$Be and pep are of approximately the same amplitude,
see also Table 1 taken from \cite{10}. We neglect here the
contributions from hep- and $^{17}F$-neutrinos.

\begin{center}
Table 1. The capture rates of solar neutrinos on chlorine, gallium
and lithium according to solar model BP2000 [10] and to the
results of SNO experiment [1,2] (for boron neutrinos).
\end{center}

\begin{tabular}{|c|c|c|c|c|}
\hline Source&Flux(10$^{10}$
sm$^{-2}$sec$^{-1}$)&\multicolumn{3}{|c|}{Capture Rate (SNU)}\\
\cline{3-5}&&chlorine&gallium&lithium\\ \hline
Pp&5.95($\pm$1\%)&-&69.7&-\\
Pep&1.40x10$^{-2}$($\pm$1.5\%)&0.22&2.8&9.2\\
Hep&9.3x10$^{-7}$&0.04&0.1&0.1\\
$^7$Be&4.77x10$^{-1}$($\pm$10\%)&1.15&34.2&9.1\\
$^8$B&1.75x10$^{-4}$($\pm$8\%)&2.0&4.2&6.8\\
$^{13}$N&5.48x10$^{-2}$(+21/-17\%)&0.09&3.4&2.3\\
$^{15}$O&4.8x10$^{-2}$(+25/-19\%)&0.33&5.5&11.8\\
$^{17}$F&5.63x10$^{-4}$($\pm$25\%)&0.0&0.1&0.1\\ \hline
Total&&3.83&124&39.4\\ \hline
\end{tabular}

\vskip 0.1in

These attractive features of lithium detector were the reason why
since the early phase of solar neutrino research lithium detector
was considered as an important element of the program of the full
neutrino spectroscopy of the Sun \cite{11,12}. Two monochromatic
lines of approximately equal intensity is a unique case for
lithium target. The flux of pep-neutrinos from the Sun is
approximately 35 times lower than the flux of $^7$Be-neutrinos
\cite{10}. So any detector sensitive to both lines will see the
effect from pep-line as a small admixture to the one from very
intensive line of $^7$Be-neutrinos. But this is not true for the
lithium detector. It happens because in the laboratory conditions
the $^7$Be line will not produce $^7$Be on lithium since the
reaction of $^7$Be production is reverse to electron capture by
$^7$Be. If to consider electron screening in terrestrial atoms,
the energy of $^7$Be line is even lower than a threshold for
$^7$Be production. But in the Sun high temperature produces the
thermal broadening of the $^7$Be line, as it was first understood
by Domogatsky \cite{13} and later was computed with high accuracy
by Bahcall \cite{14}. Because of this some fraction of the line
with the energy higher than the threshold will produce $^7$Be and
the calculations show \cite{10} that the yield of $^7$Be by this
channel is almost equal to the yield by pep-neutrinos as one can
see from Table 1. If the long-wave vacuum oscillations are
realized which is one of the probable solutions by the presently
available data [15-20] than the pattern of the seasonal variations
in lithium detector will be very characteristic, which can be used
both for the needs of the full-spectroscopy of solar neutrinos and
for the determination of the parameters of neutrino oscillations
as it was proposed in \cite{5}.

The general idea of using the seasonal variations is based on the
connection of the shape of the curve of the seasonal variations
with the oscillation parameters. The variation of the effect for a
monochromatic source in case of vacuum oscillations is described
by the following expression:

\begin{center}
$R(E,t)$ = $\Phi \sigma (E) $(1-2$\varepsilon $cos$\pi $t)[1 - sin$^2$2$\theta $sin$^2$%
(1.9x10$^{11}\Delta $m$^2$r(t)/$E$)]
\end{center}

Here $\Phi $ is the flux of neutrinos calculated for the distance
Sun-Earth 1 a.u., $\sigma (E)$ - the neutrino capture
cross-section, the value $\Phi \sigma $ is measured in SNU \\(1
SNU is the capture rate per second in 10$^{36}$ atoms of the
target), $\varepsilon $ = 0.0165,\\ r(t) = 1 + $\varepsilon
cos\pi$t, t varies from 0 (aphelion) till 1 (perihelion), this
corresponds to
one half of a year, beginning from aphelion. The factor  $\Phi \sigma $(1-2$%
\varepsilon $cos$\pi $t) describes the seasonal variations of
1/r$^2$ which accounts for 3.3\% modulation of the solar flux. For
the continuous sources ($^{13}$N, $^{15}$O, $^8$B) the integral
was taken:

\begin{center}
$R(t)$ = $\int_{E_{TH}} \frac{\varphi (E)}{\Phi} R(E,t)dE$
\end{center}

\noindent here $\varphi (E)$ is the flux of the neutrinos per
energy interval $dE$, $E_{TH}$ = 0,86MeV - the threshold of
lithium detector. The values $\varphi (E)$ and $\sigma (E)$ were
taken from \cite{21}. The calculations were done to find the curve
of the expected seasonal variations as a function of $\Delta
$m$^2$ in the maximal mixing case. The results are presented on
Fig.2 where for each value of $\Delta $m$^2$ on the axis X three
numbers are given: the annual average of the effect, the maximal
and minimal annual values of the effect. The difference of maximal
and minimal values gives the depth of the modulation that is the
most important parameter in view of the experimental accuracy
needed. To give more definite description of the shape of the
curve it would be worth to give also the frequency of the main
mode and the phase of the variations relative to, say, aphelion.
But for our current aims three parameters will be enough. Figure 2
shows the corresponding values as a function of $\Delta $m$^2$ for
$^7$Be, pep, $^{13}$N, $^{15}$O, $^8$B and for the sum of all
these sources for lithium detector. One can see that the depth of
the modulation of the signal is very dependent on $\Delta $m$^2$.

The curves of the expected seasonal variations in lithium detector
were calculated for two sets of $\Delta $m$^2$ and
sin$^2$2$\theta$: one of 1.4x10$^{-10}$ eV$^{2}$ and
sin$^2$2$\theta$ = 0.8 corresponds to the best fit point of the
global solutions in the free flux analysis and another  one of
4.8x10$^{-10}$ eV$^{2}$ and sin$^2$2$\theta$ = 0.9 corresponds to
the best fit point of the global solutions in the SSM restricted
flux analysis [18].

\begin{figure}[!t]
\centering
\includegraphics[width=2in]{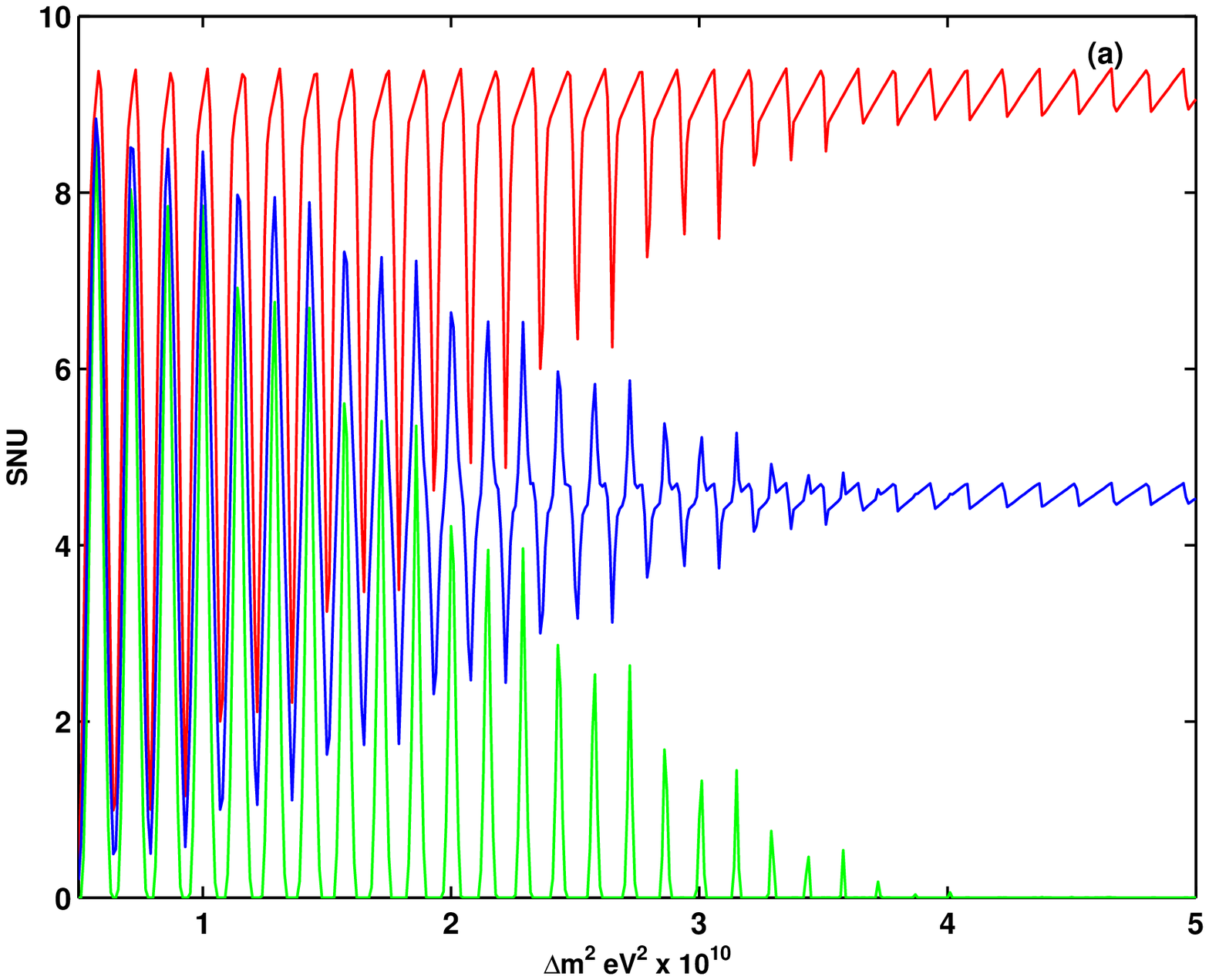}
\includegraphics[width=2in]{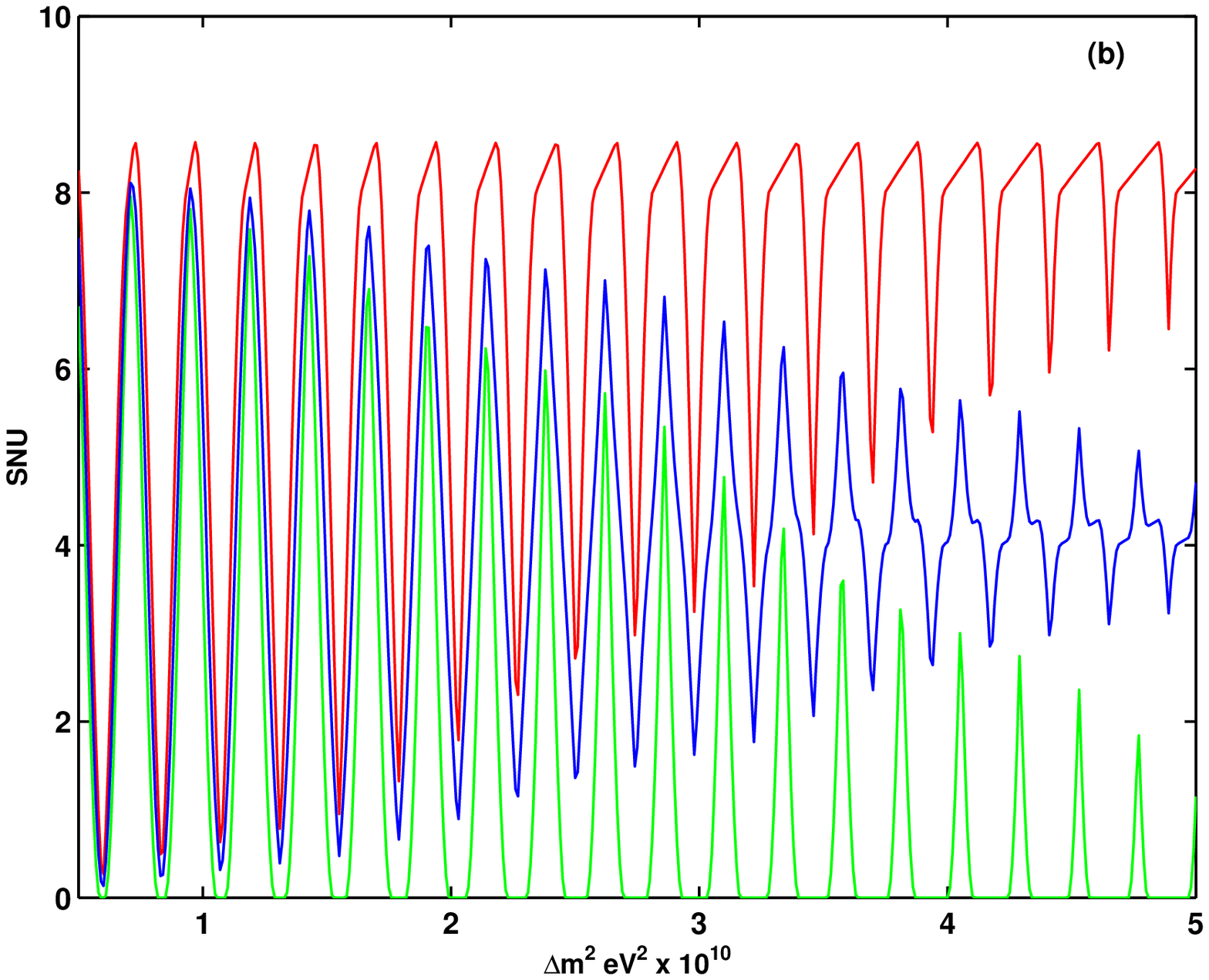}
\includegraphics[width=2in]{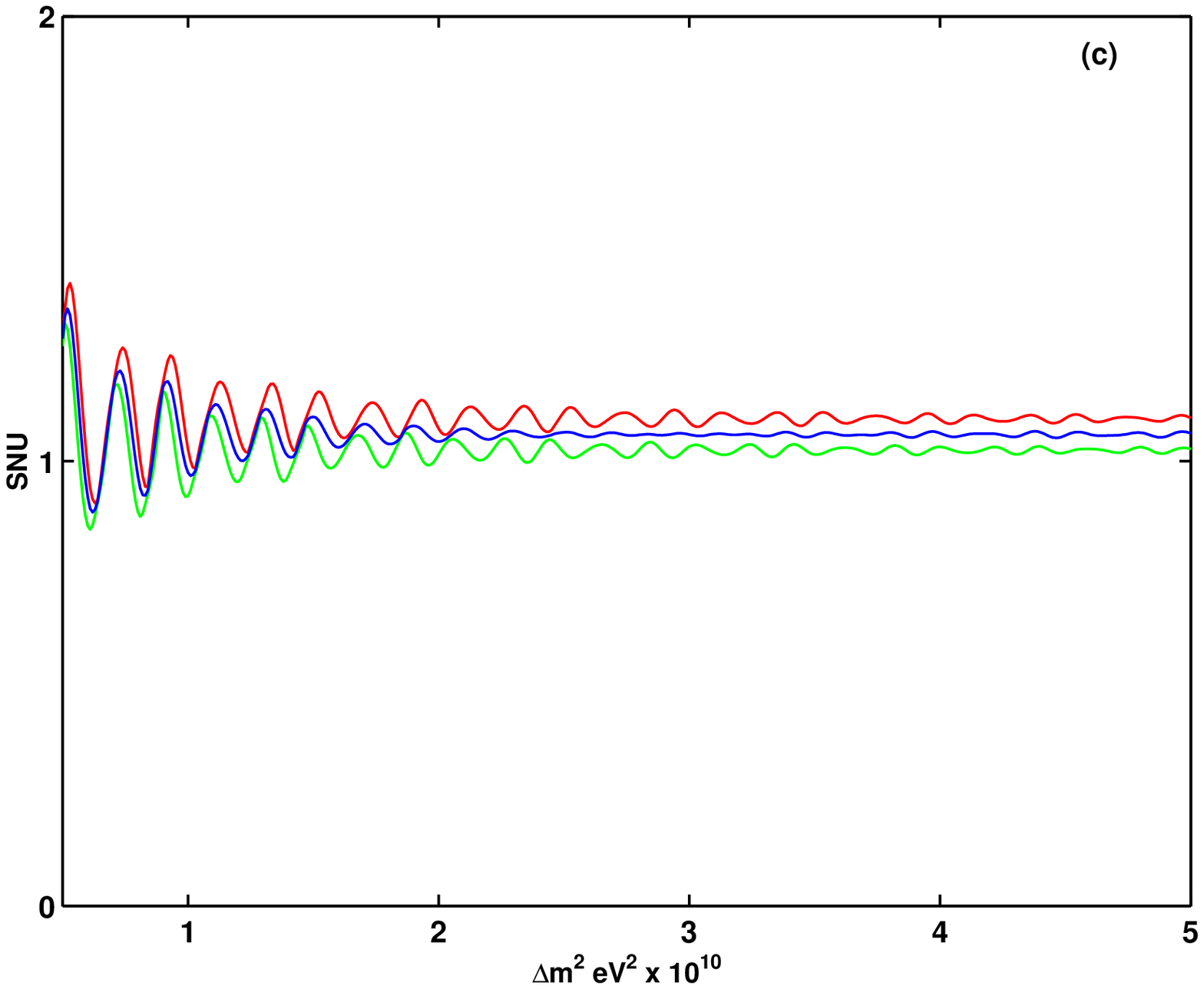}
\includegraphics[width=2in]{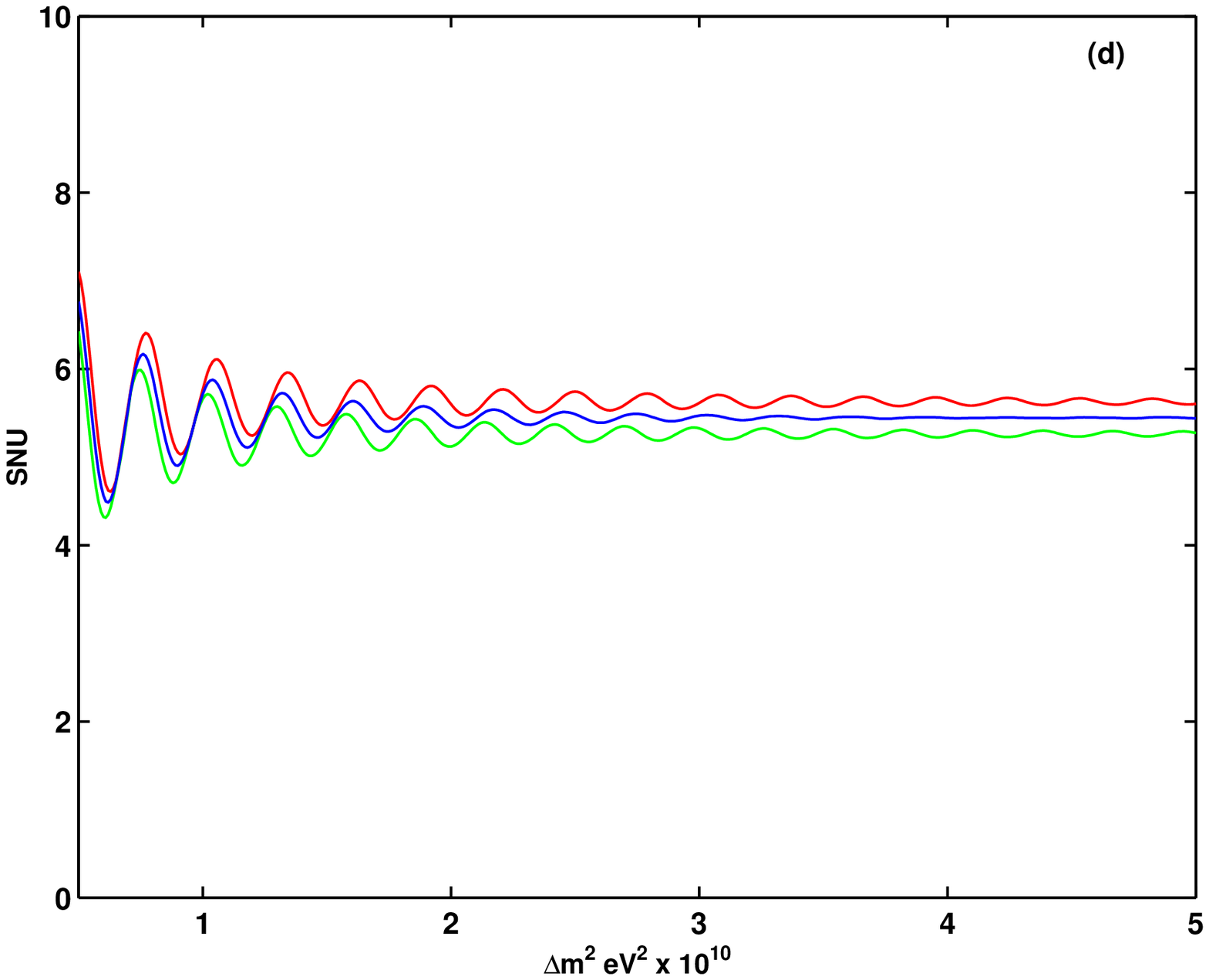}
\includegraphics[width=2in]{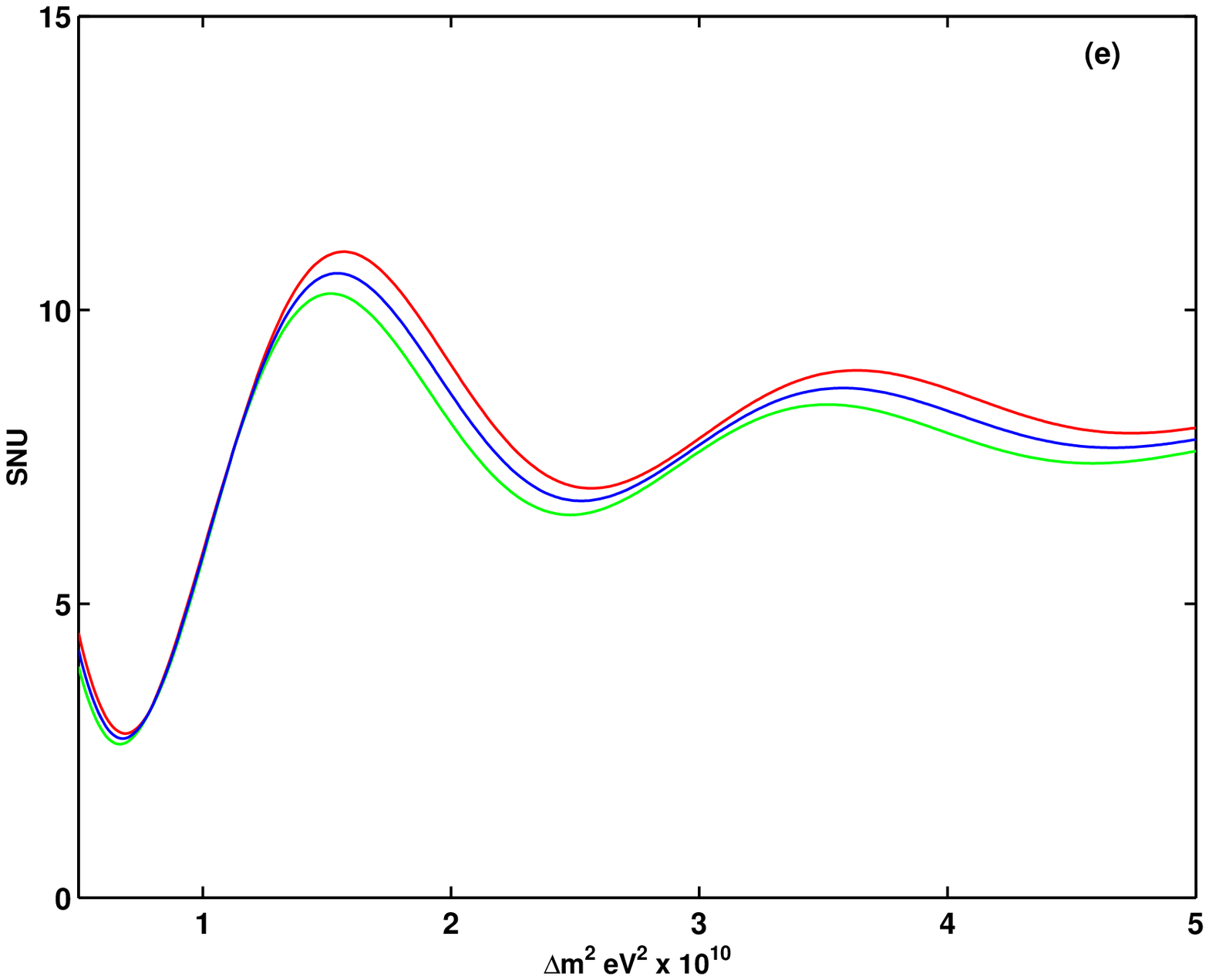}
\includegraphics[width=2in]{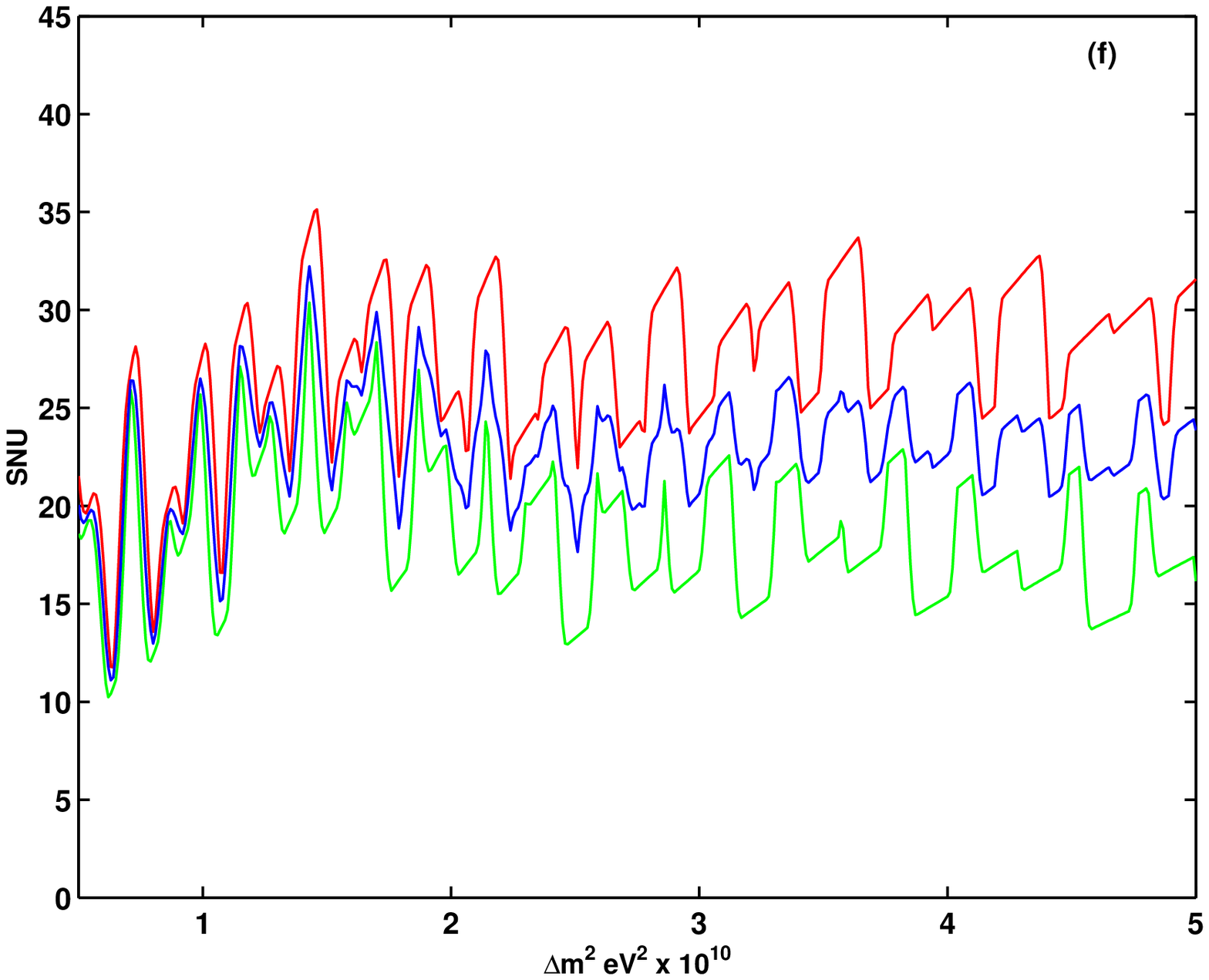}
\caption{The max (red), min (green) and average (blue) annual
values of the rates for lithium detector: (a) - $^7$Be, (b) - pep,
(c) - $^{13}$N, (d) - $^{15}$O, (e) - $^8$B, (f) - sum.}
\end{figure}

\noindent They differ drastically as one can see from Fig.3. So
these two solutions can be well resolved by lithium detector. The
evaluated mass of the lithium target adequate to perform the high
precision measurements of the seasonal variations is 100 tons.
Then the accuracy of each one-month point will be about 2.5\% for
4 years of experiment if to measure the $^7$Be activity by means
of a cryogenic microcalorimeter \cite{22,23}.

The shape of the curve of the seasonal variations is modified with
the energy of neutrinos detected in experiment. The overlap of the
allowed regions obtained by lithium detector and by the one
sensitive only to $^7$Be line will enable to reach higher
accuracy.

\begin{figure}[!t]
\centering
\includegraphics[width=3in]{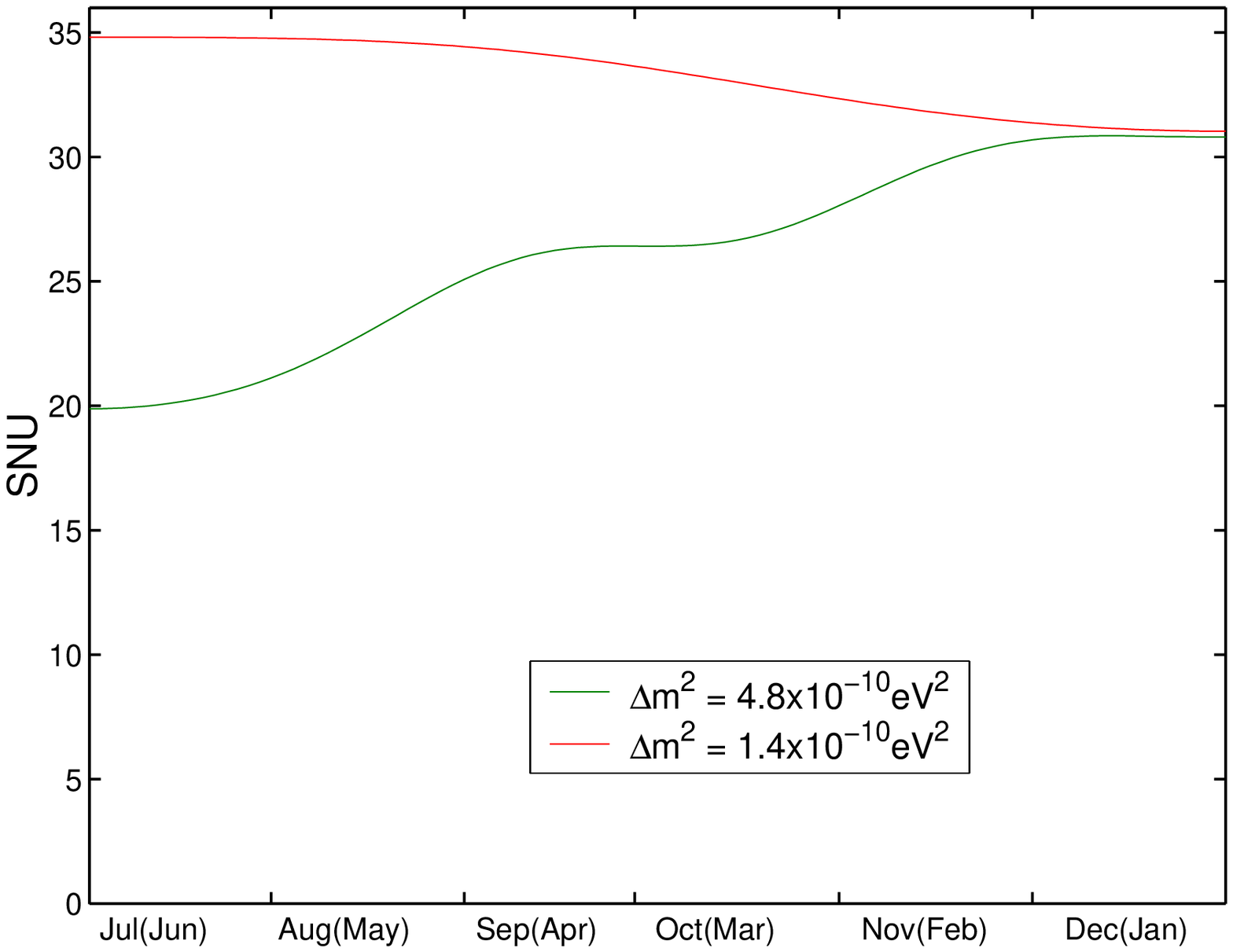}
\caption{The seasonal variations of the effect in lithium detector
for $\Delta $m$^2$=1.4x10$^{-10}$ eV$^{2}$, sin$^2$2$\theta$=0.8
and $\Delta $m$^2$=4.8x10$^{-10}$ eV$^{2}$, sin$^2$2$\theta$=0.9.}
\end{figure}

\begin{figure}[!h]
\centering
\includegraphics[width=3in]{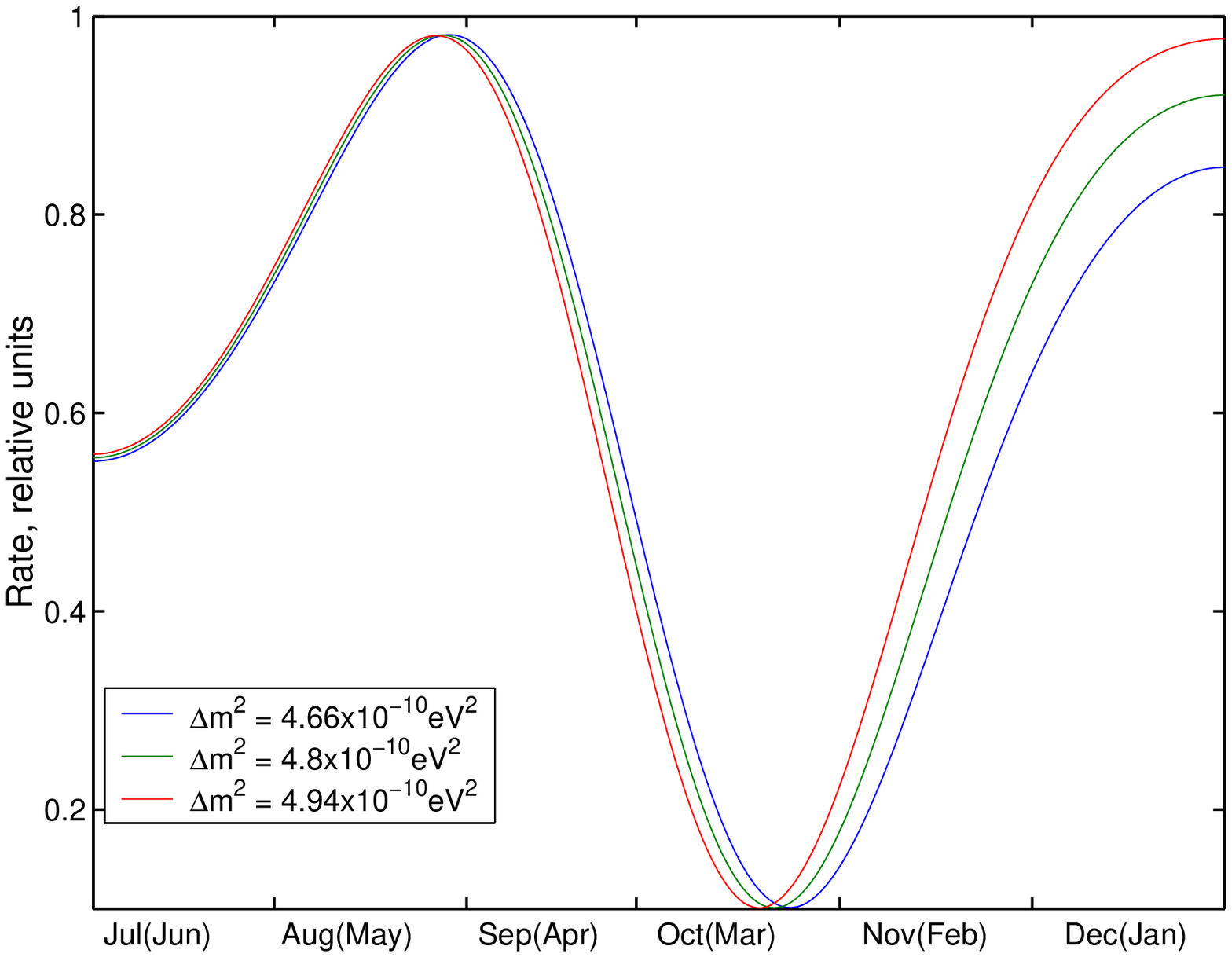}
\caption{The seasonal variations for $^7$Be line for $\Delta
m^2$=4.66x10$^{-10}eV^{2}$, $\Delta m^2$=4.8x10$^{-10}eV^{2}$ and
$\Delta m^2$=4.94x10$^{-10}eV^{2}$.}
\end{figure}

\noindent If to take, for example, three adjacent values of
$\Delta $m$^2$ on Fig.2(a) which have the same annual average and
which are close to 4.8x10$^{-10}$ eV$^{2}$ then the shapes of the
seasonal curves for the detector sensitive only to $^7$Be
neutrinos will be very alike as one can see from Fig.4. In
experiment it may be difficult to find what is the true value of
$\Delta $m$^2$ because the accuracy of the measurements may be not
sufficient to make the reliable selection between these
alternatives. For the detector sensitive to other line, pep-line,
the shapes of the curves for the same three values of $\Delta
$m$^2$ will be very different as one can see from Fig.5. It would
be much easier in this case to discriminate between these three
cases.

\begin{figure}[!t]
\centering
\includegraphics[width=2.8in]{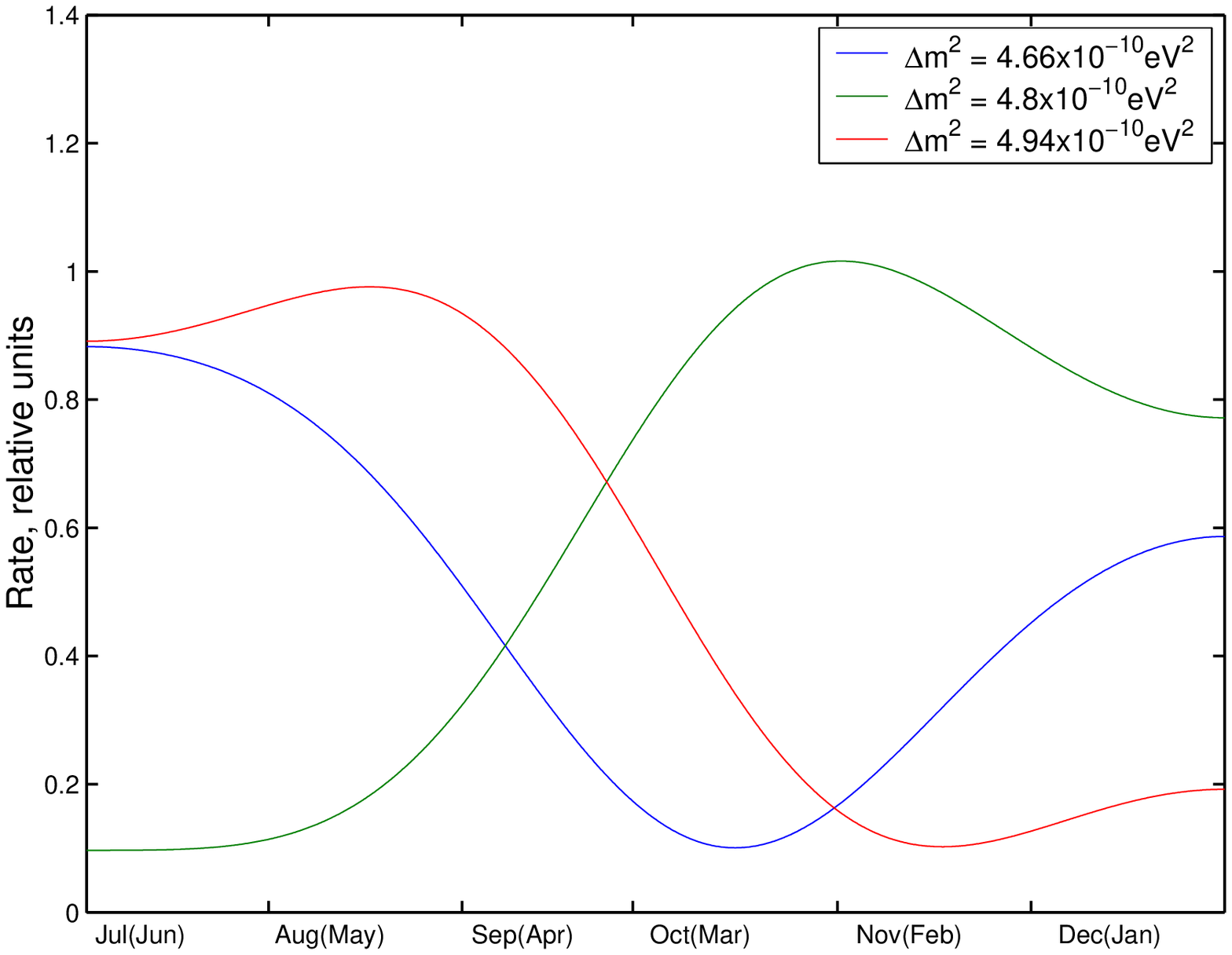}
\caption{The seasonal variation for pep line for $\Delta
m^2$=4.66x10$^{-10}$ eV$^{2}$, $\Delta m^2$=4.8x10$^{-10}$
eV$^{2}$ and $\Delta m^2$=4.94x10$^{-10}$ eV$^{2}$.}
\end{figure}

\begin{figure}[!b]
\centering
\includegraphics[width=2.8in]{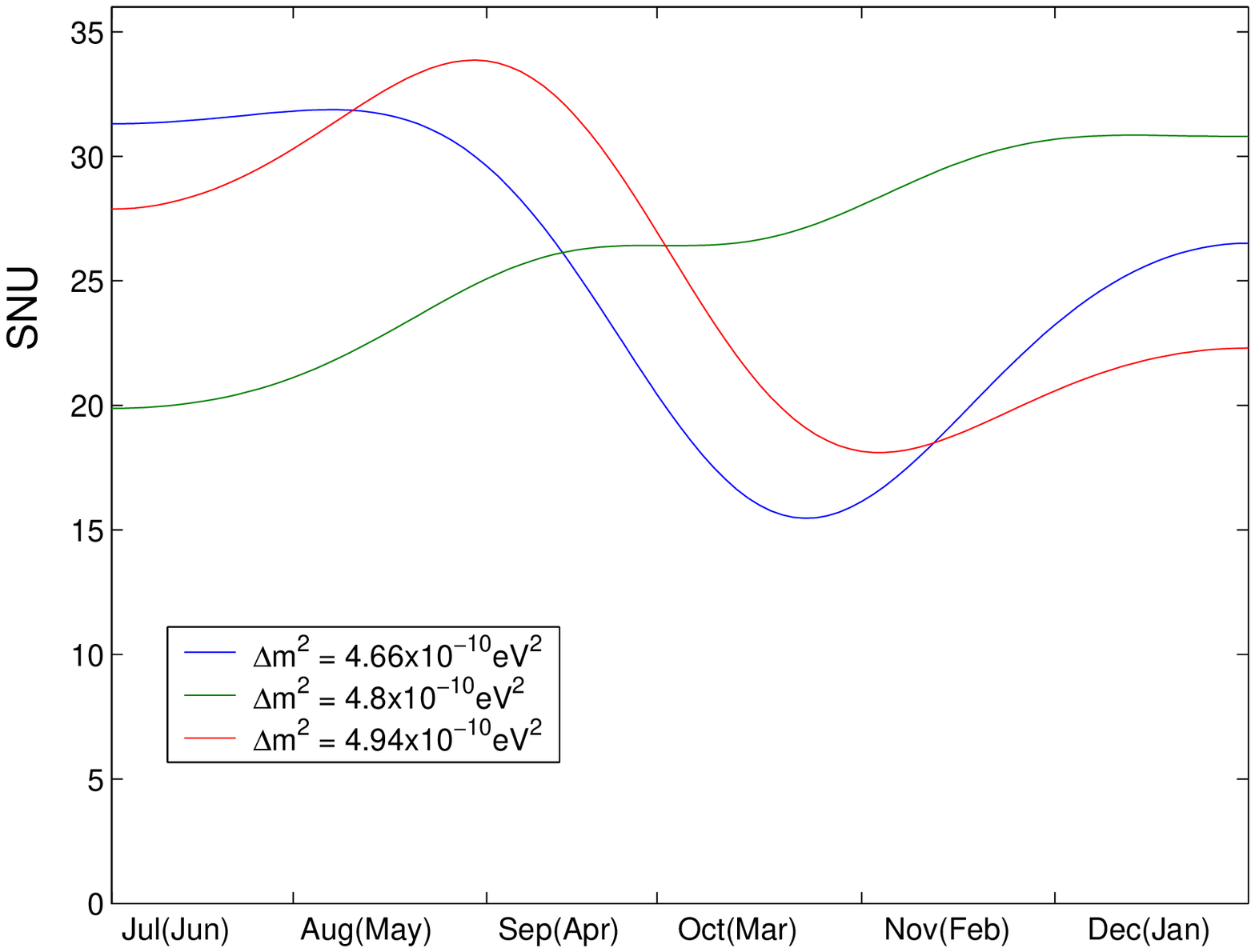}
\caption{The seasonal variations of the effect in lithium detector
for $\Delta m^2$=4.66x10$^{-10}$ eV$^{2}$, $\Delta
m^2$=4.8x10$^{-10}$ eV$^{2}$ and $\Delta m^2$=4.94x10$^{-10}$
eV$^{2}$.}
\end{figure}

The conclusion is that the comparison of the seasonal variations
obtained by $^7$Be-detector and pep-detector will enable to find
the $\Delta $m$^2$ with the very high accuracy. The lithium
detector is sensitive to several sources of neutrinos, this
creates some smearing effect. But still, the presence of two
monochromatic lines of high and approximately equal intensity
makes this detector very efficient in providing the independent
information to determine $\Delta $m$^2$. Figure 6 shows the curves
for lithium detector for the same values of $\Delta $m$^2$. One
can see the big difference of the shapes that can be efficiently
used for finding what is the true value of $\Delta $m$^2$.

The authors are grateful to G. Zatsepin and L. Bezrukov for many
fruitful discussions. This work was supported in part by the
Russian Fund of Basic Research, contract N 01-02-16167-A and by
the Leading Russian Scientific School grant N 00-15-96632.

\end{document}